\documentclass[12pt]{article}
\usepackage{graphicx}
\usepackage{color}
\newcommand{\beq}{\begin{equation}}
\newcommand{\eeq}{\end{equation}}
\newcommand{\bea}{\begin{eqnarray}}
\newcommand{\eea}{\end{eqnarray}}

\newcommand{\gsim}{\lower.7ex\hbox{$
\;\stackrel{\textstyle>}{\sim}\;$}}
\newcommand{\lsim}{\lower.7ex\hbox{$
\;\stackrel{\textstyle<}{\sim}\;$}}

\newcommand{\eod}{\end{document}}

\usepackage{epsfig}

\definecolor{verm}{rgb}{0.8,0.1,0.0}

\begin{document}
\thispagestyle{empty}
\vspace*{-22mm}

\begin{flushright}

UND-HEP-17-BIG\hspace*{.08em}02\\
Version 2.0 \\
\today \\

\end{flushright}

\vspace*{1.3mm}

\begin{center}
{\Large {\bf "The School of Athens" \\
or \\
\vspace{3mm}
"Bridge between MEP and HEP" \footnote{Contribution to the conference RADCOR 2017}}}

\vspace*{10mm}

{\em €}

\vspace*{10mm}

{\bf I.I.~Bigi$^a$} \\
\vspace{7mm}
$^a$ {\sl Department of Physics, University of Notre Dame du Lac}\\
{\sl Notre Dame, IN 46556, USA}

\vspace*{-.8mm}

{\sl email addresses: ibigi@nd.edu}

\vspace*{10mm}

{\bf Abstract}\vspace*{-1.5mm}\\
\end{center}

\noindent
This century the situation has changed: neutrino oscillations have shown that the SM is incomplete. On the other hand, 
we have not found signs of SUSY, which is seen as the best candidate for New Dynamics (ND). We should search for ND indirectly  
with accuracy in regions that are above what LHC can find it directly. General comments: 
(a) QCD is the only local quantum field theory to describe strong forces. We have to apply non-perturbative QCD on different levels to flavor dynamics 
in strange, charm \& beauty hadrons and even for top quarks. We need consistent parameterization of the CKM matrix and apply to 
weak decays of beauty hadrons with many-body final states. 
(b) It is crucial to use the Wilsonian OPE as much as possible and discuss "duality" in the worlds of quarks and hadrons. The pole mass of heavy quarks is  
{\em not} well-defined on the non-perturbative level -- i.e., it is {\em not} Borel summable in total QCD. 
(c) We need a novel team to combine the strengths of our tools from MEP and HEP.

\vspace{3mm}

\hrule

\tableofcontents

\vspace{5mm}

\hrule\vspace{5mm}

Attending the "13th International Symposium on Radiative Corrections" in Sankt (!) Gilgen (Austria) I might think I am at the wrong party. 
One talks about New Dynamics (ND) with {\em perturbative} QCD: jet productions with many hadrons -- or quarks \& gluons -- with the 
scale $\Lambda_{\rm QCD} \sim 0.1 - 0.3$ GeV. Indeed, perturbative QCD is very good tool for describing jets on the log scale:  it changes slowly. 

However, the weak decays of beauty \& charm hadrons are produced mostly with 3- \& 4-body FS. Those transitions are shaped by the impact 
of non-perturbative QCD with a scale $\bar \Lambda \sim {\cal O}(1)$ GeV, which moves fast, namely not on the log scale. 
Not surprisingly it is similar to lattice QCD.  It is affected by thresholds, resonances -- in particular for broad ones -- etc.;  
in my view I am not truly on the wrong party, as the subtitle of this conference: `Applications of Quantum Field Theory to Phenomenology'. 

\begin{figure}[h!]
\begin{center}
\includegraphics[scale=1.00]{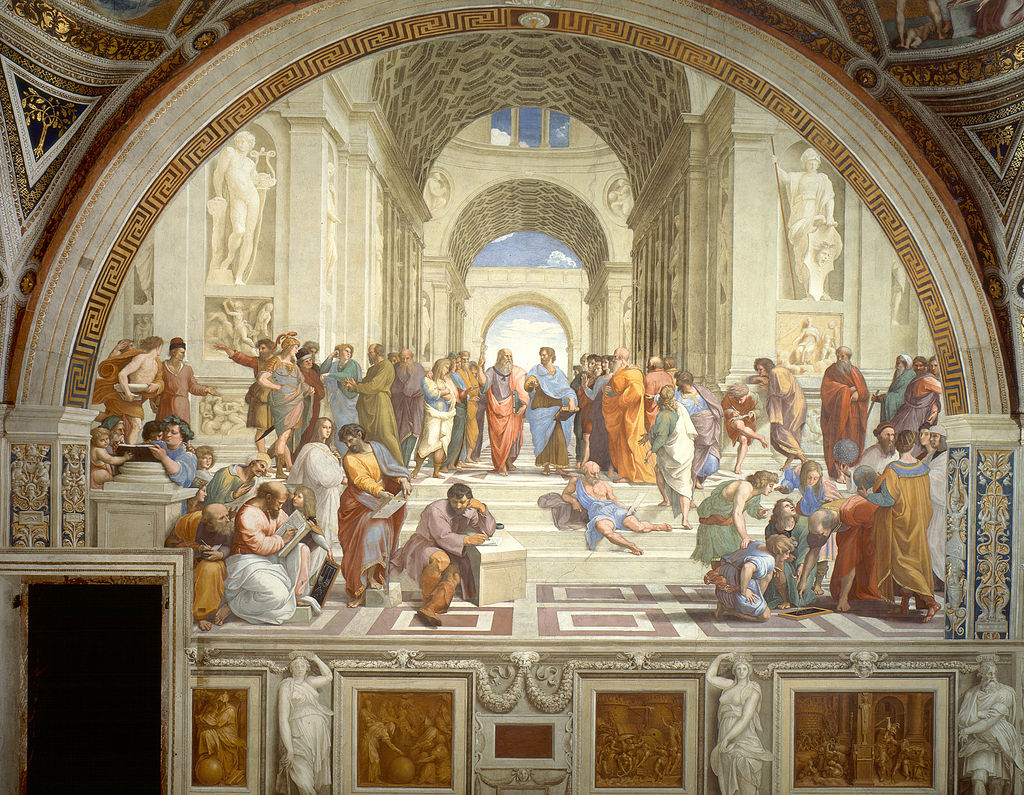}
\end{center}
\caption{The School of Athens} 
\label{fig:ATHENS}
\end{figure}

To put it in the history of fundamental dynamics: in the previous century one talks about Nuclear Physics (NP) and HEP; then flavor dynamics was 
part of HEP. Now this century the landscape has changed very much: one talks about NP and Middle Energy Physics (MEP) and HEP. 
The tools of HEP apply to jets, Higgs forces, top quarks, {\em direct} SUSY etc.; when one analyses the decays of strange, beauty \& charm 
hadrons, one uses Dalitz plots, dispersion relations etc. and the goal is to go for accuracy or even precision. We need 
New Alliance between MEP \& HEP.  One can `paint' the landscape as an analogy with the School of Athens, see Fig.\ref{fig:ATHENS}. 
In the centre one sees `Plato' pointing `up' -- thus HEP -- while `Aristotle' with one hand going forward shows much `balance' with the other hand 
-- thus MEP.

This painting can give us an analogy to the items discussed in the Sections \ref{CKM}, \ref{DUAL}, \ref{MANY}, \ref{RENORM}, \ref{ALL}, as I will say in the end (I hope). 

\section{Describing the CKM Matrix consistently}
\label{CKM}

Wolfenstein's parameterization was very smart, easily usable \& well-known. The SM with three families of quarks 
describes the CKM matrix with four parameters, namely 
$\lambda$, $A$, $\bar \rho$ \& $\bar \eta$. One uses expansion of the  
Cabibbo angle $\lambda = {\rm sin}\theta_C \simeq 0.223$, while $A$, $\bar \rho$ and $\bar \eta$ should be 
of the order of unity \cite{WOLFMAT}. 
Measured decays give $A \simeq 0.82$, but also $\bar \eta \simeq 0.35$ and $\bar \rho \simeq 0.14$, which are not close to unity; thus we have not real 
control over systematic uncertainties.

The SM produces at least the leading source of {\bf CP} violation in $K_L \to 2 \pi$ and
$B$ decays with good accuracy. Searching for ND we need even precision and to measure
the correlations with other FS's. The landscape of the CKM matrix is more subtle as pointed out
through $ {\cal O}(\lambda ^6)$ consistently  \cite{AHN}: 
\begin{eqnarray}
{\bf {\rm V}}_{\rm CKM} \simeq
\left(\footnotesize
\begin{array}{ccc}
 1 - \frac{\lambda ^2}{2} - \frac{\lambda ^4}{8} - \frac{\lambda ^6}{16}, & \lambda , &
 \bar h\lambda ^4 e^{-i\delta_{\rm QM}} , \\
 &&\\
 - \lambda + \frac{\lambda ^5}{2} f^2,  &
 1 - \frac{\lambda ^2}{2}- \frac{\lambda ^4}{8}(1+ 4f^2)
 -f \bar h \lambda^5e^{i\delta_{\rm QM}}  &
   f \lambda ^2 +  \bar h\lambda ^3 e^{-i\delta_{\rm QM}}   \\
    & +\frac{\lambda^6}{16}(4f^2 - 4 \bar h^2 -1  ) ,& -  \frac{\lambda ^5}{2} \bar h e^{-i\delta_{\rm QM}}, \\
    &&\\
 f \lambda ^3 ,&
 -f \lambda ^2 -  \bar h\lambda ^3 e^{i\delta_{\rm QM}}  &
 1 - \frac{\lambda ^4}{2} f^2 -f \bar h\lambda ^5 e^{-i\delta_{\rm QM}}  \\
 & +  \frac{\lambda ^4}{2} f + \frac{\lambda ^6}{8} f  ,
  &  -  \frac{\lambda ^6}{2}\bar h^2  \\
\end{array}
\right)
\end{eqnarray} 
with $\bar h \simeq 1.35$, $f\simeq 0.75$ \& $\delta _{\rm QM} \sim 90^{o}$ and only
expansion in $\lambda \simeq 0.223$. 
The pattern in flavor dynamics is less obvious for {\bf CP} violation in hadron decays as stated before \cite{BUZIOZ};
the situation has changed: we have to measure the correlations between four triangles, not focus only on the `golden triangle'.
Some of the important points are emphasized: 

\noindent 
(a) The maximal SM value of $S(B^0 \to J/\psi K_S)$ for indirect {\bf CP} violation is $\sim 0.74$. 

\noindent 
(b) For $S(B^0_s \to J/\psi \phi)$ is $\sim 0.03 - 0.05$. 

\noindent 
(c) The SM gives basically zero {\bf CP} value for doubly Cabibbo suppressed transitions. 

\noindent 
One has to measure accurately the correlations with several triangles.

\section{Duality: Measuring $|V_{qb}|$ with $q=c,u$}
\label{DUAL}

The item of "duality" is referred to a very complex situations, namely the connections of the worlds of hadrons vs. quark \& gluons. I will come back to the 
{\bf Sect.\ref{RENORM}}. In this Section I give short comments at the very specific case: compare the values of $|V_{cb}|$ and $|V_{ub}|$ from inclusive 
vs. exclusive semi-leptonic amplitudes. 

It seems the difference between the $|V_{cb}|_{\rm incl.}$ vs. $|V_{cb}|_{\rm excl.}$ has become smaller now based on realistic 
theoretical uncertainties, mostly due to LQCD analyses. 

On the other hand, the difference between $|V_{ub}|_{\rm incl.}$ vs. $|V_{ub}|_{\rm excl.}$ has not changed. It has been pointed out that 
the values of $|V_{ub}|_{\rm incl}$ based on the data from $B \to l \nu \pi$'s, while assuming that $B \to l \nu \bar KK ...$ are irrelevant due to a 
traditional understand duality. It is a good assumption -- but {\em local} duality does not work close to thresholds. Maybe the real $|V_{ub}|_{\rm incl}$ 
are smaller and thus solve that challenge. LHCb experiment cannot measure inclusive rates. However, it might go able to go after the rates of 
$B^+ \to l^+\nu K^+K^-$ and $B^0 \to l^+\nu K^+K^-\pi^-$ with non-zero values. Furthermore Belle II should measure values there or limits.

\section{Many-body Final States for $\Delta B \neq 0 \neq \Delta C$ Hadrons}
\label{MANY}

{\bf CP} asymmetries depend on 
\beq
|T(\bar P \to \bar a)|^2 - |T(P \to  a)|^2 = 4\sum_{a_j,a} T_{a_j,a}^{\rm resc} \; {\rm Im} T^*_a T_{a_j} \; ; 
\eeq 
with{\em out non-zero re-scattering} direct {\bf CP} asymmetries cannot happen, if there are weak phases 
\cite{1988BOOK,WOLFFSI,FSICP,TIM}. 
Even so, it is a very good hunting region for the impact of New Dynamics (ND), since they can depend only one ND amplitude. 
\begin{itemize}
\item
One expects large impact of strong re-scattering, and the data of suppressed $B \to 3$ mesons have shown that.

\item
In particular, there is a `fuzzy' difference between broken U-spin symmetry -- $s \rightleftharpoons d$ -- and broken V-spin symmetry -- 
$s \rightleftharpoons u$. 

\item
Probing FS in non-leptonic decays with two hadrons (including narrow resonances) is not trivial to measure {\bf CP} violations; on the 
other hand one gets `just' numbers. Three- \& four-body FS are described by two \& more dimensional plots. There is a price: lots of 
work for experimenters and theorists. There is also a prize: to find the existence of ND and also its features.

\end{itemize}

\subsection{Probing Dalitz plot for $B^{\pm}$}

The data of CKM suppressed $B^+$ decays show no surprising rates for $B^+ \to K^+\pi^-\pi^+$ \& $B^+ \to K^+K^-K^+$ and 
$B^+ \to \pi^+\pi^-\pi^+$ \& $B^+ \to \pi^+K^-K^+$. 

LHCb data from run-1 show averaged direct {\bf CP} asymmetries \cite{LHCb028}:
\bea
\nonumber
\Delta A_{CP}(B^{\pm} \to K^{\pm} \pi^+\pi^-) &=&
+0.032 \pm 0.008_{\rm stat} \pm 0.004_{\rm syst}
\pm 0.007_{\psi K^{\pm}}
\\
\Delta A_{CP}(B^{\pm} \to K^{\pm} K^+K^-) &=&
- 0.043 \pm 0.009_{\rm stat} \pm 0.003_{\rm syst}
\pm 0.007_{\psi K^{\pm}}
\label{SUPP2}
\eea
with 2.8 $\sigma$ \& 3.7 $\sigma$ from zero. Based on our experience with the impact of
penguin diagrams on the best measured $B^0 \to K^+\pi^-$, the sizes of these averaged asymmetries are not surprising; however it does not mean that we
could really predict them. It is very
interesting that they come with opposite signs due to {\bf CPT} invariance.

LHCb data show {\em regional} {\bf CP} asymmetries \cite{LHCb028}:
\bea
\nonumber
A_{CP}(B^{\pm} \to K^{\pm} \pi^+\pi^-)|_{\rm regional} &=& + 0.678 \pm 0.078_{\rm stat}
\pm 0.032_{\rm syst}
\pm 0.007_{\psi K^{\pm}}
\\
A_{CP}(B^{\pm} \to K^{\pm} K^+K^-)|_{\rm regional} &=& - 0.226 \pm 0.020_{\rm stat}
\pm 0.004_{\rm syst} \pm 0.007_{\psi K^{\pm}} \; .
\label{SUPP4}
\eea
"Regional" {\bf CP} asymmetries are defined by the LHCb collaboration: positive asymmetry at low
$m_{\pi ^+\pi ^-}$ just below $m_{\rho^0}$; negative asymmetry both at low and high $m_{K^+K^-}$ values. One should note again the opposite signs in Eqs.(\ref{SUPP4}).
It is not surprising that
"regional" asymmetries are very different from averaged ones. Even when one uses states only from
the SM -- $SU(3)_C \times SU(2)_L \times U(1)$ -- one expects that; it shows the
{\em impact of re-scattering}
due to $SU(3)_C$ (actually $SU(3)_C \times$QED) in general. Of course, our community needs more data,
but that is not enough. There are important questions
and/or statements:
\begin{itemize}
\item
How do we {\em define} regional asymmetries and probe them on the experimental and theoretical sides?

\item
Can it show the impact of broad resonances like $f_0(500)/\sigma$ and $K^*(800)/\kappa$?

\item
Again, the best fitted analyses often do not give us the best understanding of the underlying
fundental dynamics.

\end{itemize}

LHCb data from the run-1 show {\em larger} averaged {\bf CP} asymmetries as discussed above in Eqs.(\ref{SUPP2})
(again with the opposite signs):
\bea
\nonumber
\Delta A_{CP}(B^{\pm} \to \pi^{\pm}  \pi^+\pi^-) &=&
+0.117 \pm 0.021_{\rm stat} \pm 0.009_{\rm syst}
\pm 0.007_{\psi K^{\pm}}
\\
\Delta A_{CP}(B^{\pm} \to \pi^{\pm} K^+K^-) &=&
- 0.141 \pm 0.040_{\rm stat} \pm 0.018_{\rm syst}
\pm 0.007_{\psi K^{\pm}}  \; .
\label{SUPP6}
\eea
It is interesting already with the averaged ones, since $b \Longrightarrow d$ penguin diagrams are more suppressed than $b\Longrightarrow s$ ones.
Again {\bf CP} asymmetries focus on small regions in the Dalitz plots \cite{LHCb028}.
\bea
\nonumber
\Delta A_{CP}(B^{\pm} \to \pi^{\pm} \pi^+\pi^-)|_{\rm regional} &=&
+0.584 \pm 0.082_{\rm stat} \pm 0.027_{\rm syst}
\pm 0.007_{\psi K^{\pm}}
\\
\Delta A_{CP}(B^{\pm} \to \pi^{\pm} K^+K^-)|_{\rm regional} &=&
- 0.648 \pm 0.070_{\rm stat} \pm 0.013_{\rm syst}
\pm 0.007_{\psi K^{\pm}} \; .
\label{SUPP8}
\eea
Again, it should be noted also the signs in Eqs.(\ref{SUPP6}) \& Eqs.(\ref{SUPP8}).
Does it show the impact of
broad scalar resonances like $f_0(500)/\sigma$ and/or $K^*(800)/\kappa$?

First one analyzes the data using model-independent techniques \cite{REIS}, compares them and discuss the results --
but that is not the end of our `traveling'. Well-known tools like dispersion relations are `waiting' to apply -- but we have to do it 
with some `judgement'. 

\subsection{Probing for {\bf CP} violation in the decays of $\Lambda_b^0$}

At the ICHEP2016 conference in Chicago it was said that LHCb data show evidence for {\bf CP} asymmetry in $\Lambda_b^0 \to p \pi^- \pi^+\pi^-$.
It is discussed in \cite{LHCBBARCP} with some details. In $pp$ collisions one gets numbers of $\Lambda_b^0$ vs. $\bar \Lambda_b^0$ due to
{\em production} asymmetries. Therefore one focuses first on {\bf T-}odd moments. LHCb measured the angle between two planes: one is formed by the momenta of
$p$ \& $\pi^-_{\rm fast}$, while the other one with the momenta of $\pi^+$ \& $\pi^-_{\rm slow}$. They found evidence for {\bf CP} asymmetry on the level of
3.3 $\sigma$ based in its run-1 of 3 fb$^{-1}$. Actually the angle was measured between one plane defined in the rest frame of $[\vec p \times \vec \pi^-_{\rm fast}]$, 
while the other one $[\vec \pi^+ \times \vec \pi^-_{\rm slow}]$. The present data can give us more information about the underlying dynamics by measuring 
the angle between different planes, namely with one plane defined by $[\vec p \times \vec \pi^-_{\rm slow}]$, while the other one $[\vec \pi^+ \times \vec \pi^-_{\rm fast}]$. 
Those should be affected by different broad resonances, thresholds etc.

The data are very interesting for several reasons:
 \begin{itemize}
\item
Maybe {\bf CP} asymmetry was found in a decay of a baryon for the first time (except `our existence'); it is for a beauty baryon. 

\item
It is one example that many-body FS are not a background for the information our
community got it from two-body FS.

\item
The plot given at the ICHEP2016 shows the strength of regional {\bf T} asymmetry around 20 $\times 10^{-2}$. Very interesting, but we cannot claim to
understand the underlying dynamics -- yet! Furthermore in the world of quarks \& gluons one looks at
CKM penguin of $b \to d$, where one expects less than for $b \to s$. LHCb data already shown similar lessons
for {\bf CP} asymmetries in $B^+ \to \pi^+\pi^+\pi^-/\pi^+K^+K^-$ vs. $B^+ \to K^+\pi^+\pi^-/K^+K^+K^-$, see 
Eqs.(\ref{SUPP2},\ref{SUPP4},\ref{SUPP6},\ref{SUPP8}) just above.

\item
LHCb collaboration did not get enough data from run-1 to probe $\Lambda_b^0 \to p \pi^-K^+K^-$ \& $\bar \Lambda_b^0 \to \bar p \pi^-K^+K^-$.
It will change very `soon'.

\end{itemize}
There are another very important comments: LHCb collab. can measure rates and {\bf CP}
"regional" asymmetries in $\Lambda_b^0 \to p K^-\pi^+ \pi^-$ and $\Lambda_b^0 \to p K^-K^+ K^-$
`soon' -- and has no competition from other experiments. First we have to discuss
$\Lambda_b^0 \to p \pi^-\pi^+\pi^-$ \& $\Lambda_b^0 \to p \pi^-K^+K^-$ and
$\Lambda_b^0 \to p K^-\pi^+\pi^-$ \& $\Lambda_b^0 \to p K^-K^+K^-$. Will they follow the
same `landscape' for $B^+ \to \pi^+\pi^+\pi^-/\pi^+K^+K^-$ vs.
$B^+ \to K^+\pi^+\pi^-/K^+K^+K^-$ as discussed above qualitatively or not? So say it
with different words: will they show the strengths of `penguin diagrams' in $\Lambda_b^0$ decays or not?
The situations are similar for beauty mesons and baryons or only on the qualitatively way?

\section{Deal with Renormalons}
\label{RENORM}

Dyson pointed out in his famous 1952 paper "Divergences of Perturbation Theory in QED" that amplitudes can{\em not} be convergent. 
Later it was realized perturbative series in a QFT are {\em factorially divergent} like $Z=\sum_k C_k \alpha^k k^{b-1}A^{-k}k!$ 
with $k \gg 1$ is the number of loops, $C_k$'s are numerical coefficients of order one, and $b$ \& $A$ are numbers. 
It is traced back to the factorially large number of multi-loop Feynman diagrams. The features responsible for the renormalon factorial divergence is 
the logarithmic running of the effective coupling constant. 

Instead of {\em asymptotic} series one can introduce a Borel transform
\beq
B_Z = \sum_k C_k \alpha^k k^{b-1}A^{-k}  \; ;
\eeq
the singularity of $B_Z (\alpha)$ closest to the origin of the $\alpha$ plain is at a distance $A$, and thus $B_Z (\alpha)$ is {\em convergent}. 
One recovers the original function $Z$ by 
\beq
Z(\alpha) = \int_0^{\infty} dt \; e^{-t}B_Z (\alpha t)
\label{ZALL}
\eeq
The integral representation is well-defined {\em provided} that $B_Z(\alpha)$ has {\em no} singularities on the real positive semi-axis in the complex $\alpha$ plane. 
That is not a problem for QED.

{\em If} $B_Z(\alpha)$ has a singularity on the real positive semi-axis -- like coefficients $C_k$ are all positive or all negative -- the integrated in the 
Eq.(\ref{ZALL}) become {\em ambiguous}. This ambiguity is of the order of $e^{-A/\alpha}$; more information is needed from the underlying dynamics. 
The question comes from QCD with 
\beq
\alpha_S (Q^2) \simeq \frac{\alpha_S(\mu^2)}{1- \frac{\beta_0 \alpha_S(\mu^2)}{4\pi} {\rm log}(\mu^2/Q^2)} =  
\frac{\alpha_S(\mu^2)}{1+ \frac{\beta_0 \alpha_S(\mu^2)}{4\pi} {\rm log}(Q^2/\mu^2)} \; , \; \beta_0 = 11  - \frac{2}{3}N_f \; ; 
\label{STRONGCOUPL}
\eeq
the energy scale $\mu$ is used to calibrate $\alpha_S (Q^2)$. The good side is: at large scales the strong couplings go down to zero with $Q^2/\mu^2$ (on the log scale) -- 
i.e. "asymptotic freedom". 

On the other hand, there is a true challenge. With $\mu^2 \gg Q^2$  $\alpha_S (Q^2)$ gets larger and larger; thus 
QCD gives us true strong forces at low scales. First one might say it goes to infinite, but that is too naive.  
One has to stop at $\mu \sim 1$ GeV based on perturbative QCD. 

It was pointed out first in 1994 that the pole mass is {\em not} well-defined at the non-perturbative level \cite{RENORM2}. 
Furthermore a rather powerful renormalon-based tool was suggested 
for evaluating the corresponding non-perturbative contribution \cite{SHIFMAN2013}. Pole mass is sensitive to large distance dynamics, although this fact is not obvious in 
perturbative calculations. IR contributions lead to an {\em intrinsic uncertainty} in the pole mass of order $\Lambda$ -- i.e., a $\Lambda/m_Q$ power correction. 
It comes from the factorial growth of the high order terms in the $\alpha_S$ expansion corresponding to a singularity residing at the $2\pi/\beta_0$ in the Borel plane. 
Thus one cannot say it is a correction. 

Actually, there are two renormalon-based tools, namely ultraviolate (UV) and infrared (IR) dynamics  
\footnote{"All animals are equal, but some animals are more equal than others!" G. Orwell, `Animal Farm'.}.  One has to include non-perturbative QCD  
with IR one. Those give contribution to $b$ quark mass numerically \cite{BSU97}:
\bea
\nonumber
m_b^{\rm pole} &=&  m_b( 1 \, {\rm GeV}) +\delta m_{\rm pert}(\leq 1\, {\rm GeV})\simeq \\
&\simeq& 4.55 \, {\rm GeV} + 0.25 \, {\rm GeV} + 0.22 \, {\rm GeV} + 0.38 \, {\rm GeV} + 1\, {\rm GeV} + 3.3\, {\rm GeV}  ... \; ,
\label{MBPOLE}
\eea
where $\delta m_{\rm pert}(\leq 1\, {\rm GeV})$ is the perturbative series taking account of the loop momenta down to zero. 
\begin{figure}[h!]
\begin{center}
\includegraphics[width=10cm]{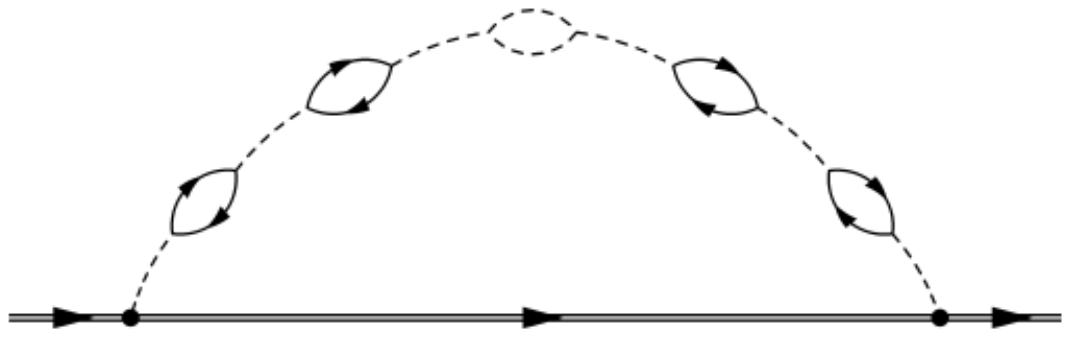}
\end{center}
 \vspace{-11cm}
\caption{ Perturbative diagrams leading to the IR renormalon
uncertainty in $m_Q^{\rm pole}$ of order $\bar \Lambda$.
The number of bubble insertions in the gluon propagator
is arbitrary. The horizontal line at the bottom is the heavy quark Green's function.}
\label{fig5}
\end{figure}

Indeed, top quarks decays before they have produce top hadrons. Still they carry unbroken color symmetry and thus find partners with color 
to produce hadrons with color zero in the FS.

\section{Summary and a New Alliance for the Future}
\label{ALL}


We have to proceed in steps:
\begin{enumerate}
\item
First we use models. 

\item
Then we use model-independent analyses. 

\item
Those are not the final step(s). 
Often best fitted analyses do not give us the best information about the underlying dynamics. How to do that? 
We have theoretical tools with a good record like dispersion relations and/or partial wave analyses; they are `waiting' -- it only 
need work with judgements and tests it with correlations!

\end{enumerate}
In the previous century one had talked about fundamental physics: Nuclear Physics at low energies, while HEP at high energies. 
Flavor dynamics had been discussed at HEP. 

In this century for sure one thinks (or should) about Nuclear Physics and MEP and HEP. Probing jets, Higgs \& top quarks dynamics 
and direct SUSY is the `job' for HEP still again. However, the landscape is more complex with many interconnected parts: decays of strange/beauty/charm 
hadrons, where tools applied to Dalitz plots with dispersion relations etc. We have to go for accuracy and even precision to find the impact of ND. 
To make true progress, it is crucial to connect the world of hadrons, where MEP applies (with a better choice of word "Hadro-dynamics"), 
with the world of HEP mostly of quarks \& gluons; it is highly non-trivial. 
Still I was not totally on the wrong party, when one looks at the sub-title of the conference: `Application of Quantum Field Theory to Phenomenology'. 

\begin{figure}[h!]
\begin{center}
\includegraphics[scale=1.50]{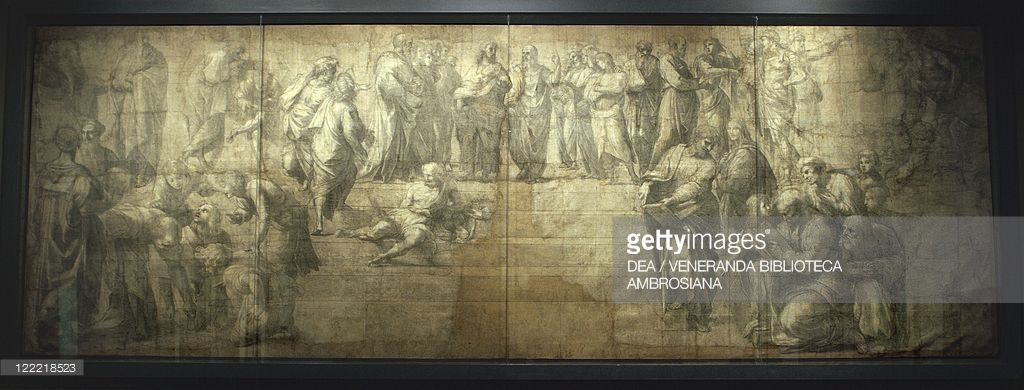}
\end{center}
\caption{Draft of the `School of Athens'} 
\label{fig:DRAFT2}
\end{figure}

In the Fig.\ref{fig:ATHENS} you had seen a painting of the 
`School of Athens'; to be realistic, I show the reader a draft of the `School of Athens' with `Plato' = theory and `Aristotle' = experiment in the center, see Fig.\ref{fig:DRAFT2}. 


\vspace{0.5cm}

{\bf Acknowledgments:} This work was supported by the NSF under the grant number PHY-1520966. I have enjoyed this conference 
in such a wonderful landscape with amazing flavor in different ways; likewise for the excellent organ concert at the Basilica Sankt Michael Mondsee. 
I want to thank to the organizers -- in particular Andre Hoang \& Carsten Schneider. 
\vspace{4mm}



\begin{thebibliography}{99}

\bibitem{WOLFMAT}
L. Wolfenstein, {\em Phys.Rev.Lett.} {\bf 51} (1983) 1945. 

\bibitem{AHN} 
Y.H. Ahn, H-Y. Cheng, S. Oh, {\em Phys.Lett.} {\bf B703} (2011) 571; arXiv: 1106.0935. 

\bibitem{BUZIOZ}
I.I. Bigi, arXiv:1306.6014 [hep-ph], talk given at FPCP2013"Flavor Physics and CP Violation 2013". 

\bibitem{1988BOOK}
I.I. Bigi, V.A. Khoze, N.G. Uraltsev, A.I. Sanda, p. 175-248 in "CP Violation", C. Jarlskog (Editor), 
World Scientific (1988).

\bibitem{WOLFFSI}
L. Wolfenstein, {\em Phys.Rev.} {\bf D43} (1991) 151.

\bibitem{FSICP}
N.G. Uraltsev, Proceedings of DPF-92, arXiv:hep-ph/9212233. 

\bibitem{TIM}
I.I. Bigi, {\em Frontiers of Physics} {\bf 10}, \#3 (June 2015) 101203; arXiv:1503.07719 [hep-ph]. 

\bibitem{LHCb028}
R. Aaij {\em et al.}, LHCb collab., {\em Phys.Rev.Lett.} {\bf 111} (2013) 101801;
R. Aaij {\em et al.}, LHCb collab., {\em Phys.Rev.Lett.} {\bf 112} (2014) 011801.

\bibitem{REIS}
I. Bediaga {\em et al.}, {\em Phys.Rev.} {\bf D89} (2014) 074024; arXiv:1401.3310 [hep-ph].

 \bibitem{LHCBBARCP}
R. Aaij {\em et al.}, LHCb collaboration, nature physics, published online: 30 January 2017; arXiv: 1609.05216 [hep-ex]. 


\bibitem{RENORM2}
I.I. Bigi, M. Shifman, N.G. Uraltsev, A. Vainshtein, {\em Phys.Rev.} {\bf D50} (1994) 2234; 
M. Beneke, V. Braun, {\em Nucl.Phys.} {\bf B426} (1994) 301.


\bibitem{SHIFMAN2013}
M. Shifman, "New and Old About Renormalons", arXiv:1310.1966 [hep-th];
contribution to the book: "QCD and Heavy Quarks -- In Memoriam Nikolai Uraltsev", World Scientific Publishing Co., 2015; it shows the progress since 1994; 
M. Shifman, "Quark-hadron duality", Boris Ioffe Festschrift, At the Frontier of Particle Physics, Singapore, World Scientific Publishing Co., 2001; arXiv: hep-ph/0009131.

\bibitem{BSU97}
I. Bigi, M. Shifman, N. Uraltsev, {\em Annu.Rev.Nucl.Sci.} {\bf 47} (1997) 591; arXiv: hep-ph/9703290.











 

















\end{thebibliography}
\end{document}